# MULTI-SOURCE WIDEBAND DOA ESTIMATION METHOD BY FREQUENCY FOCUSING AND ERROR WEIGHTING

*Jing Zhou, Changchun Bao*[*]

Speech and Audio Signal Processing Laboratory, Faculty of Information Technology, Beijing University of Technology, Beijing 100124, China.
zhoujing@emails.bjut.edu.cn; baochch@bjut.edu.cn

## ABSTRACT

In this paper, a new multi-source wideband direction of arrival (MSW-DOA) estimation method is proposed for the signal with non-uniform distribution using the sub-array of uniform linear array. Different from conventional methods, based on the free far-field model, the proposed method mainly makes two contributions. One is that the sub-array decomposition is adopted to improve the accuracy of MSW-DOA estimation by minimizing the weighted error, and the other one is that the frequency focusing procedure is optimized according to the presence probability of sound sources for reducing the influence of the sub-bands with low signal to noise ratio (SNR). Simulation results show that the proposed method can effectively improve the performance of wideband DOA estimation in the case of multiple sound sources.

**Index Terms:** microphone array, direction of arrival, frequency focusing, subspace decomposition

## 1. Introduction

Multi-source wideband direction of arrival (MSW-DOA) estimation is a research hotspot in the fields of radar, sonar, wireless communication, speech signal processing, etc. [1-10]. An accurate direction of arrival (DOA) estimation method can provide guarantee for the subsequent processing and applications, such as in beamforming [6]. The conventional wideband DOA estimation methods are based on the Fourier transform [11-13], which transform the signal from time domain to frequency domain, and further design the wideband DOA estimation methods according to the narrowband DOA estimation methods [9,11-13]. For example, the incoherent signal subspace (ISS) method proposed by Su [14] used the mean of the estimated DOAs of all sub-bands as the result of wideband DOA estimation. However, the signal may not be uniform distribution within the bandwidth, which may cause a great error of DOA estimation in the sub-band with lower signal-to-noise ratio (SNR), and ISS method is not suitable for the coherent signals processing [8,9]. Therefore, Wang [15] proposed the coherent signal subspace (CSS) method by frequency focusing, which can reduce the coherence by smoothing within sub-bands [16,17], and make the covariance matrix with full rank [7-9].

The core of CSS method is the procedure of frequency focusing [9,17]. Conventional frequency focusing methods include signal subspace transformation (SST) [18], rotational signal sub-space (RSS) [16], two-sided correlation transform-ation (TCT) [19], etc. Those methods are based on the premise that the focusing matrix is an unitary matrix and the noise is Gaussian white noise [5,9]. In addition, most focusing methods need to estimate a preliminary DOA, but it may cause a large error while inaccuracy [2~9]. Thus, Ma [8] proposed the focusing signal subspace (FSS) method, which mainly focused on the point of do not have to estimate the preliminary DOA [4,9,11,12]. Although FSS method seems very effective to MSW-DOA estimation, there are still many drawbacks in practical application, such as the noise may not be Gaussian white noise, source signals are non-uniform distribution within bandwidth, the signal subspace is very sensitive to the error of focusing model, etc [3,4,8,11]. Therefore, to overcome the above limitations, a new FSS based method is proposed in this paper for improving the performance of MSW-DOA estimation.

## 2. Signal model and problem description

Considering an uniform linear array (ULA) composed of $M+B$ microphones, the distance between the microphones is $d$. Based on the free far-field model, the ULA can be divided into $B+1$ successive sub-arrays, and the microphone number of each sub-array is $M$, as shown in figure 1. Based on this particular division, we can ensure same performance for each sub-array, and the DOA of each sound source can be regarded as the same for each sub-array. Assuming there are $Q$ sound sources contained in acoustic field, the time domain model of the output observed signal $x_b(t)$ of the $b^{th}$ sub-array can be expressed as follows [4,8]:

$$\boldsymbol{x}_b(t) = \sum_{q=1}^{Q} \boldsymbol{s}_{b,q}(t) + \boldsymbol{n}_b(t) \quad (1)$$

where $\boldsymbol{s}_{b,q}(t)$ and $\boldsymbol{n}_b(t)$ are the $q^{th}$ sound source and noise relate to the $b^{th}$ sub-array at time $t$, respectively. $b=1,2,\ldots,B+1$.

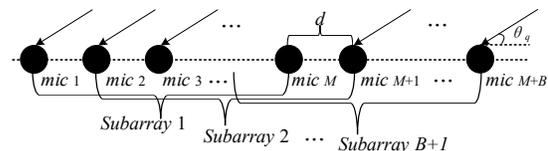

Figure 1: *Uniform linear array and sub-arrays*

Taking the center microphone of each sub-array as the reference microphone, and performing $K$-point short time Fourier transform (STFT) on Eq. (1), the observed signal at the $k^{th}$ sub-band of the $l^{th}$ frame is obtained as follows:

$$\boldsymbol{X}_b(f_k, l) = \sum_{q=1}^{Q} \boldsymbol{a}_b(f_k, \theta_q) S_q(f_k, l) + \boldsymbol{N}_b(f_k, l) \quad (2)$$

where $X_b(f_k,l)$, $S_q(f_k,l)$ and $N_b(f_k,l)$ are the STFT results of $x_b(t)$, $s_q(t)$ and $n_b(t)$, respectively. $s_q(t)$ is the signal of the $q^{th}$ sound source. $f_k$ is the frequency of the $k^{th}$ sub-band, $\theta_q$ is the incident angle of the $q^{th}$ sound source. $\boldsymbol{a}_b(f_k,\theta_q)$ is the steering vector [4-12] of the $q^{th}$ sound source at the $k^{th}$ sub-band relates to the $b^{th}$ sub-array. The index *l* will be omitted hereinafter for coherent expression.

According to Eq. (2), the observed signal covariance matrix (OSCM) of the $b^{th}$ sub-array $\mathbf{R}_b(f_k)$ can be estimated [4,9]. Moreover, the influence of frequency factor can be eliminated by focusing on each sub-band based on the focusing matrix [4,8,9,16]. In CSS and FSS, the construction of focusing matrix and the selection of reference frequency are two key points, which will directly affect the performance of DOA estimation [9,12]. Suppose the focusing matrix relates to the $b^{th}$ sub-array is $\mathbf{C}_b(f_k)$, the focusing covariance matrix (FCM) of $k^{th}$ sub-band can be estimated as:

$$\mathbf{R}_{b,k}(f_0) \approx \frac{1}{J}\sum_J \left[\mathbf{C}_b(f_k)\mathbf{X}_b(f_k)\right]\left[\mathbf{C}_b(f_k)\mathbf{X}_b(f_k)\right]^H \quad (3)$$

where $J$ is the number of snapshots, that is, $J$ frames are used to estimate OSCM. $f_0$ is the reference frequency, superscript "H" represents the conjugate transpose operation.

According to Eq. (3), the smoothed focusing covariance matrix (SFCM) can be expressed as:

$$\overline{\mathbf{R}}_b(f_0) = \frac{1}{K}\sum_{k=1}^{K}\mathbf{R}_{b,k}(f_0) \quad (4)$$

Performing eigenvalue decomposition on Eq. (4), the noise subspace $\mathbf{U}_{N,b}(f_0)$ is obtained by the eigenvectors relate to the last $M$-$P$ smaller eigenvalues [7-9], $P$ is the number of the significantly large eigenvalue values. According to the orthogonality principle of subspace decomposition, the estimated DOAs are obtained by searching the maximum values of MUSIC spectrum [9,10]:

$$\phi_b(\theta) = \frac{1}{\mathbf{a}_b^H(f_0,\theta)\mathbf{U}_{N,b}(f_0)\mathbf{U}_{N,b}^H(f_0)\mathbf{a}_b(f_0,\theta)} \quad (5)$$

where $\mathbf{a}_b(f_0,\theta)$ is the steering vector with incident angle $\theta$ at reference frequency $f_0$, $\theta \in [0°, 180°]$.

The conventional CSS method needs a preliminary DOA to construct $\mathbf{C}_b(f_k)$ [8,12,20]. However, when there is a large error in the preliminary DOA, it is easy to result the DOA estimation in failures [8,9,12,13,20]. Therefore, some subspace methods without the preliminary DOA estimation are proposed in [4,8,9], such as FSS. Since the non-uniform distribution of sound sources and noise within bandwidth, the SNR of some sub-bands is lower, or even does not contain the components of the sound sources. The above case is easy lead to the diffusion from the noise subspace to the signal subspace [12,16,19], that means those sub-bands will cause the orthogonality between the steering vectors of the sound sources and noise subspace $\mathbf{U}_{N,b}(f_0)$ get worse. In addition, there are deviations in the focusing process of FSS, which also lead to the decline in the accuracy of DOA estimation, especially in the case of focusing within the whole bandwidth [4].

## 3. The proposed method

### 3.1. Sub-band selection

In order to reduce the diffusion from noise subspace to signal subspace, the effective sub-bands in the snapshots need to be selected before solving the focusing matrix $\mathbf{C}_b(f_k)$. Since the non-uniform distribution of the sound sources, the SFCM cannot be calculated by taking few sub-bands, which is easy to cause some sound sources to be missed or miscalculation. Therefore, we consider identifying the effective frequency bins in the snapshots through the sound source presence probability (SSPP) [21-25], thus the effective sub-bands can be selected. In this paper, a hidden Markov model (HMM) based method proposed in [24] is used to estimate the SSPP. Suppose the estimated SSPP of the $k^{th}$ sub-band relates to the $b^{th}$ sub-array is $p_b(k)$ [24], $p_b(k)$ is between 0~1. Generally, the sub-bands which are suitable for DOA estimation can be adjudicated by an appropriate threshold function according to the standard deviation $\sigma_{b,k}$ and mean $\mu_{b,k}$ of $p_b(k)$ as follows:

$$p_b(k) = \begin{cases} 0, \sigma_{b,k}^2 \leq T_1 \, \& \, \mu_{b,k} \leq T_2 \\ 1, others \end{cases} \quad (6)$$

where $p_b(k)=0$ or 1 indicate the sub-band is discarded or retained, respectively. $T_1$ and $T_2$ are two thresholds.

### 3.2. Frequency focusing and weighted smoothing

The research in [4] shows that the deviation of focusing process is smaller within the bandwidth of 500Hz, so that we divided the bandwidth into several sub-bandwidths, and focusing in each sub-bandwidth to reduce the deviation of focusing process. Suppose the bandwidth is divided into $W$ sub-bandwidths, and the frequency of the selected sub-band in the $w^{th}$ sub-bandwidth is $f_{w,k}$, so that the OSCM of the $w^{th}$ sub-bandwidth $\mathbf{R}_b(f_{w,k})$ can be calculated. Performing singular value decomposition (SVD) on $\mathbf{R}_b(f_{w,k})$, and the signal subspace $\mathbf{V}_{b,S}(f_{w,k})$ can be obtained according to the largest $P_w$ singular values [4,9].

By minimizing the deviation of focusing, the conditional constraint can be express as follows [8]:

$$\begin{cases} \arg\min_{f_{w,k}} \sum_{k=1}^{K_w} \left\|\mathbf{V}_{S,b}(f_{w,0}) - \mathbf{C}_b(f_{w,k})\mathbf{V}_{S,b}(f_{w,k})\right\|_2^2 \\ \text{subject to } \mathbf{C}_b(f_{w,k})\mathbf{C}_b^H(f_{w,k}) = 1 \end{cases} \quad (7)$$

where $f_{w,0}$ is the reference frequency, $\mathbf{V}_{S,b}(f_{w,0})$ is the signal subspace relates to the sub-band of $f_{w,0}$, $\mathbf{C}_b(f_{w,k})$ is the focusing matrix relates to $f_{w,k}$, $K_w$ is the sub-band number of $w^{th}$ sub-bandwidth, $\|.\|_2$ represents the 2-norm operation.

The optimal $f_{w,0}$ can be obtained by searching $f_{w,k}$ on Eq. (7), and the focusing matrix can be constructed as follows:

$$\mathbf{C}_b(f_{w,k}) = \mathbf{V}_{S,b}(f_{w,0})\mathbf{V}_{S,b}^H(f_{w,k}) \quad (8)$$

Substituting Eq. (8) into Eq. (3), $\mathbf{R}_{b,k}(f_{w,0})$ can be obtained. In addition, in order to reduce the influence of the sub-bands with lower SNR, the smoothing process of the FCM are weighted by the mean value of SSPP of each sub-band as follows:

$$\mathbf{R}_{b,w}(f_0) = \sum_{k=1}^{K_w} \overline{p}_b(f_{w,k})\mathbf{R}_{b,k}(f_{w,0}) \quad (9)$$

$$\overline{p}_b(f_{w,k}) = \frac{1}{J}\sum_J p_{b,w}(k) \quad (10)$$

where $p_{b,w}(k)$ and $\overline{p}_b(f_{w,k})$ are the SSPP and mean value of SSPP relate to the $k^{th}$ sub-band of the $w^{th}$ sub-bandwidth, respectively. Further, the MUSIC spectrum can be obtained by Eq. (5), and the results of the estimated DOAs of each sub-array can be obtained by searching the maximum values and averaging on $W$ sub-bandwidths.

### 3.3. Error weighting and iteration

Since the orthogonality between the noise subspace and the steering vector will get worse in the case of non-Gaussian noise and lower SNR, we consider to correct the estimated DOA by the method of weighted error minimization [26]. Suppose the DOA set of the $l^{th}$ frame estimated by $B+1$ sub-arrays is the matrix $\boldsymbol{\theta}$, which can be expressed as follows:

$$\boldsymbol{\theta} = [\boldsymbol{\theta}_1, \boldsymbol{\theta}_2, ..., \boldsymbol{\theta}_b, ..., \boldsymbol{\theta}_{B+1}] = \begin{bmatrix} \theta_{1,1} & \theta_{1,2} & \cdots & \theta_{1,b} & \cdots & \theta_{1,B+1} \\ \theta_{2,1} & \theta_{2,2} & \cdots & \theta_{2,b} & \cdots & \theta_{2,B+1} \\ \vdots & \vdots & \ddots & \vdots & \ddots & \vdots \\ \theta_{q,1} & \theta_{q,2} & \cdots & \theta_{q,b} & \cdots & \theta_{q,B+1} \\ \vdots & \vdots & \ddots & \vdots & \ddots & \vdots \\ \theta_{Q,1} & \theta_{Q,2} & \cdots & \theta_{Q,b} & \cdots & \theta_{Q,B+1} \end{bmatrix} \quad (11)$$

where $\boldsymbol{\theta}_b$ is the vector of the estimated DOA set of the $Q$ sound sources relates to the $b^{th}$ sub-array. $\theta_{q,b}$ is the estimated DOA of the $q^{th}$ sound source relates to the $b^{th}$ sub-array.

In free far-field, it can be regarded that each sound source has the same incident angle relates to the $B+1$ sub-arrays, thus the error of $\theta_{q,b}$ can be defined as:

$$E_{q,b} = |\theta_{q,b} - \theta_{q,r}| \quad (12)$$

so the overall error of the estimated DOAs can be defined as:

$$E_{\text{overall}} = \sum_{b=1}^{B+1} \sum_{q=1}^{Q} E_{q,b} \quad (13)$$

where $\theta_{q,r}$ is the real DOA of the $q^{th}$ sound source.

Assuming $E_{q,b}$ obeys the zero-mean Gaussian distribution, so that the variance of $E_{q,b}$ can be defined as:

$$\sigma_q^2 = \frac{1}{B+1} \sum_{b=1}^{B+1} E_{q,b}^2 \quad (14)$$

thus, the probability distribution function of $\theta_{q,b}$ can be expressed as follows:

$$F_{q,b} = \frac{1}{\sqrt{2\pi}\sigma_q} \exp\left(-\frac{E_{q,b}^2}{2\sigma_q^2}\right) \quad (15)$$

and the weight of DOA error can be modeled as:

$$\omega_{q,b} = F_{q,b} \Big/ \sum_{b=1}^{B+1} F_{q,b} \quad (16)$$

Therefore, the objective function of minimizing the weighted error can be expressed as:

$$\theta_{q,c} \Leftarrow \arg\min_{\theta_{q,c}} \sum_{b=1}^{B+1} \omega_{q,b} |\theta_{q,b} - \theta_{q,c}| \quad (17)$$

the corrected DOA $\theta_{q,c}$ can be obtained by searching in $[0°\sim180°]$. By repeating the procedure of Eq. (17), the corrected DOAs of the $Q$ sound sources can be obtained.

From Eq. (17), we can find that the zero-mean Gaussian distribution can reduce the sensitivity to excessive values of $E_{q,b}$, so that in theory, this error weighting method could reduce the overall error $E_{\text{overall}}$. Moreover, the smaller the distribution range of DOA error, the more consistent with the assumption of zero-mean. Therefore, the iterative of DOA correction is used to reduce the deviation of the assumption of Gaussian distribution. Let the initial value of $\omega_{q,b}$ be $1/(B+1)$, so that the initial corrected DOAs can be obtained by Eq. (17).

When the procedure of initial correction is done, we can establish a new focusing objective function, and recalculate the reference frequency. The process of the $\eta^{th}$ iteration of the $w^{th}$ sub-bandwidth relates to the $b^{th}$ sub-array can be expressed as follows:

$$\min_{f_{w,0}^{\langle\eta\rangle}} \sum_{k=1}^{K_w} \left\| \mathbf{A}_b\left(f_{w,0}^{\langle\eta\rangle}, \boldsymbol{\theta}_c^{\langle\eta\rangle}\right) - \mathbf{C}_b\left(f_{w,k}, \boldsymbol{\theta}_c^{\langle\eta\rangle}\right) \mathbf{A}_b\left(f_{w,k}, \boldsymbol{\theta}_c^{\langle\eta\rangle}\right) \right\| \quad (18)$$

$$\mathbf{C}_b\left(f_{w,k}, \boldsymbol{\theta}_c^{\langle\eta\rangle}\right) = \mathbf{G}_{b,1}\left(f_{w,0}^{\langle\eta\rangle}, \boldsymbol{\theta}_c^{\langle\eta\rangle}\right) \mathbf{G}_{b,2}^{-1}\left(f_{w,k}, \boldsymbol{\theta}_c^{\langle\eta\rangle}\right) \quad (19)$$

$$\mathbf{G}_{b,1}\left(f_{w,0}^{\langle\eta\rangle}, \boldsymbol{\theta}_c^{\langle\eta\rangle}\right) = \left[\mathbf{A}_b\left(f_{w,0}^{\langle\eta\rangle}, \boldsymbol{\theta}_c^{\langle\eta\rangle}\right), \mathbf{H}\right] \quad (20)$$

$$\mathbf{G}_{b,2}\left(f_{w,k}, \boldsymbol{\theta}_c^{\langle\eta\rangle}\right) = \left[\mathbf{A}_b\left(f_{w,k}, \boldsymbol{\theta}_c^{\langle\eta\rangle}\right), \mathbf{H}\right] \quad (21)$$

where superscript "$\langle\eta\rangle$" indicates the $\eta^{th}$ iteration. $\boldsymbol{\theta}_c^{\langle\eta\rangle} = [\theta_{1,c}^{\langle\eta\rangle}, \theta_{2,c}^{\langle\eta\rangle}, ..., \theta_{q,c}^{\langle\eta\rangle}, ..., \theta_{Q,c}^{\langle\eta\rangle}]$ is the vector of corrected DOA in the $\eta^{th}$ iteration, $f_{w,0}^{\langle\eta\rangle}$ is the reference frequency in the $\eta^{th}$ iteration. $\mathbf{A}_b(f_{w,k}, \boldsymbol{\theta}_c^{\langle\eta\rangle}) = [\mathbf{a}_b(f_{w,k}, \theta_{1,c}^{\langle\eta\rangle}), ..., \mathbf{a}_b(f_{w,k}, \theta_{q,c}^{\langle\eta\rangle}), ..., \mathbf{a}_b(f_{w,k}, \theta_{Q,c}^{\langle\eta\rangle})]$ is a combinatorial matrix of the steering vectors of the $Q$ sources relates to the $b^{th}$ sub-array, $\mathbf{A}_b(f_{w,0}^{\langle\eta\rangle}, \boldsymbol{\theta}_c^{\langle\eta\rangle})$ is the similar combinatorial matrix under reference frequency $f_{w,0}^{\langle\eta\rangle}$. $\mathbf{C}_b(f_{w,k}, \boldsymbol{\theta}_c^{\langle\eta\rangle})$ is the focusing matrix by the $\eta^{th}$ iteration, $\mathbf{H}^T = [\mathbf{0}_{(M-Q)\times Q}, \mathbf{I}_{(M-Q)\times(M-Q)}]$ is the supplementary directional response matrix [15].

Performing the process of cross iteration through DOA correction and focusing signal subspace until $E_{\text{overall}}$ in Eq. (13) and $f_{w,0}^{\langle\eta\rangle}$ in Eq. (18) are stably convergent, thus the final estimated DOAs are obtained.

## 4. Simulation

In the simulation, the microphone numbers of ULA and each sub-array is set to 16 and 6, respectively. The number of the sub-array is 11, the distance $d$ between microphones is set to 0.02m and the acoustic speed $c$ is set to 340m/s. The number of sound sources in the acoustic field going up to 4 at the same time is also considered. TIMIT corpus is used for the simulation, and 200 utterances are randomly selected for each sound source. The noise is babble noise, and the multi-channel noisy speech data are generated by a simulator given in [27]. Taking speech source 1 as the reference signal, the signal to interference ratio (SIR) for other speech sources is set as 0dB, the SNR are set to -10dB, -5dB, 0dB, 5dB, 10dB, 15dB and 20dB. The real incident angles of speech sources are randomly generated from $[0°\sim180°]$, and the incident angle distances of the adjacent speech sources are constrained to be greater than 5°. The evaluation measures used in this paper are root mean square error (RMSE) under different SNR and snapshot size [4,7,8,19], and the number of cross iterations is 25. The threshold values of $T_1$ and $T_2$ in Eq. (6) are set to 0.05 and 0.1, respectively. The proposed method with the process of sub-band selection and smoothing by SSPP (SSPP-FSS), and the proposed SSPP-FSS with weighted error minimization (SSPP-WEM-FSS). Those two methods are compared with single sub-band ISS (SSB-ISS) [7], FSS [8], ISS [14], and FSS within direct-path dominance (FSS-DPD) [4].

Figure 2 shows the results of MUSIC spectrum under different number of the sound sources. Among them, the number of the snapshots is 41, and the SNR is 5dB. In figure 2, it can be found that SSB-ISS cannot detect the peak values relate to the DOAs of the sound sources, FSS cannot detect the DOAs in the case of multiple sound sources, ISS shows a stable performance variation with the number of the sound sources, FSS-DPD has obtained a better MUSIC spectrum with sharp peaks and low valleys. In addition, figure 2 also shows that the proposed SSPP-FSS obtained a smoother MUSIC spectrum than FSS-DPD, and the proposed SSPP-WEM-FSS obtained a MUSIC spectrum with sharper peaks and lower valleys. Moreover, the peaks of SSPP-WEM-FSS are closer to real DOAs and without pseudo peak, which indicate that it has the best performance.

In addition, figure 3 shows some details of the cross iterations. From figure 3(a) and figure 3(b), we can find that as the number of iteration increases, the overall error $E_{overall}$ gradually decreases, the reference frequency $f_{w,0}$ of the 5$^{th}$ sub-bandwidth gradually become stable, and $E_{overall}$ converges through 20$^{th}$ to 25$^{th}$ iterations. Those results verify the feasibility of the proposed method to reduce DOA errors by error weighting and iteration.

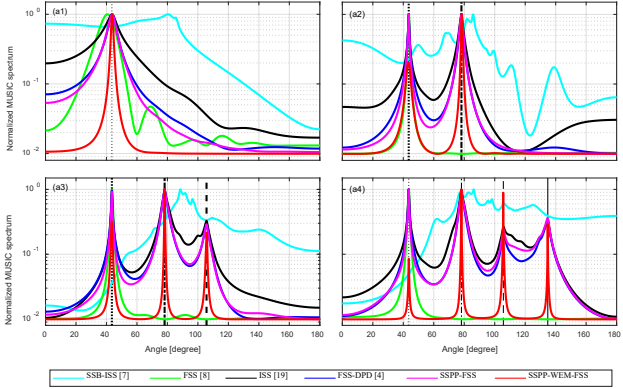

Figure 2: *Normalized MUSIC spectrum in the case of different number of sound sources with babble noise of 5dB, (a1) single source, (a2) two sources, (a3) three sources, (a4) four sources.*

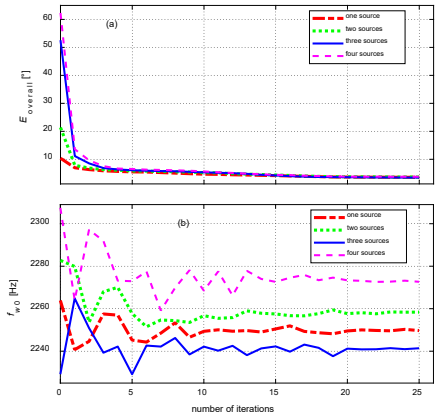

Figure 3: *The results of the cross iterations, (a) $E_{overall}$ versus the number of iterations, (b) $f_{w,0}$ of the 5th sub-bandwidth versus the number of iterations.*

Figure 4 shows the RMSE of DOA error versus the SNR of babble noise with different number of the sound sources. The number of the snapshots is 41. It should be noted that if the DOA of sound source is not detected, the DOA error is directly regarded as 10°. Figure 4 shows that in the case of different number of the sound sources, the proposed SSPP-WEM-FSS obtained the minimum RMSE values under different SNR of babble noise, which verified that the proposed SSPP-WEM-FSS method has higher DOA estimation accuracy.

In addition, figure 5 shows the RMSE of DOA error versus the number of the snapshots, and we can find that comparison with other methods, the proposed SSPP-WEM-FSS method obtained a smaller RMSE under different number of the snapshots.

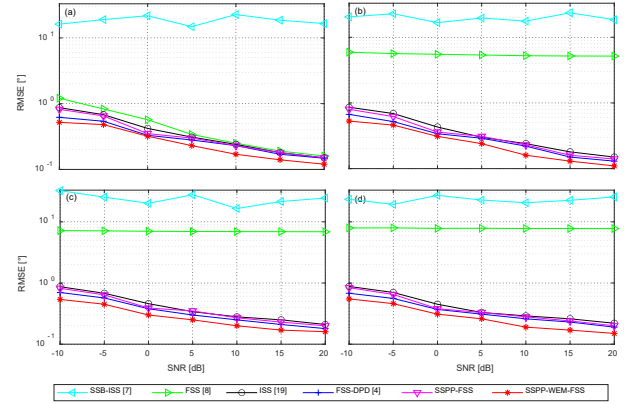

Figure 4: *RMSE of DOA error versus SNR of babble noise, (a) single source, (b) two sources, (c) three sources, (d) four sources.*

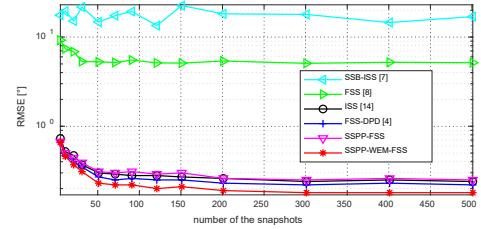

Figure 5: *RMSE of DOA error versus the number of the snapshots with 5dB babble noise, and the number of the sound sources is 2.*

## 5. Conclusions

This paper presented a new method for MSW-DOA estimation by frequency focusing and error weighting. In this method, based on the principle of FSS without preliminary DOA estimation, SSPP is used to select the effective sub-bands, and its mean is used to smooth the FCM, which reduced the occurrence of pseudo peaks. Moreover, the DOA of each sound source was corrected by minimizing the weighting DOA error of the sub-arrays and iterating the process of frequency focusing, thus more accurate MSW-DOA estimation was achieved. Simulation results verified that the presented method has smaller RMSE of DOA in the case of different SNR, different number of the sound sources and different number of the snapshots. In addition, this method can also be extended to the array of other structure, such as the planar array, which is constructed by ULA.

## 6. Acknowledgment

This work was supported by the National Natural Science Foundation of China (Grant No.61831019).